# Nanomechanical resonance captures pre-melting transition in DNA unravelling

Keren Jiang[1], Faheem Khan[1], Javix Thomas[1], Arindam Phani[1,2*] and Thomas Thundat[1,2*]

[1] Department of Chemical and Materials Engineering, University of Alberta.
Edmonton, Alberta, T6G 2V4, Canada.

[2] Department of Chemical and Biological Engineering, University at Buffalo.
Buffalo, New York, 14260-4200, USA.

[*] Corresponding authors. E-mail: aphani@buffalo.edu and tgthunda@buffalo.edu

**Abstract:**

A double-stranded DNA unravels thermally through intermediate denatured bubble segments. Intrinsically, fluctuations ensue at the bubble boundaries from non-equilibrium (NE) energy exchanges with the environment. However, such details gets obscured by large population kinetics at the macroscale, associating equilibrium pathway to the unravelling landscape. In this work, we capture evidence of fluctuation energetics with picoliter samples in a microfluidic cantilever. We exploit nanomechanical resonance to measure the NE energy exchanges through dissipation, revealing a crucial pre-melting transition at $T \sim 42^0 C$. This signifies that unravelling possibly proceeds via intermediate collapsed-bubble conformations releasing energy, sufficient to unbind bubble ends, assisting further unbinding. Fluctuation theorem explains the observations opening further avenues to investigate fluctuation kinetics in other biological phenomena that also proceed through similar NE energetics.

**One sentence Summary:**

Resonance of a microfluidic cantilever captures intermediate collapsed-bubble stages in DNA unravelling at biologically relevant $T \sim 42^0 C$.

Biological outcomes are described from an equilibrium state. However, nothing remains in absolute equilibrium. Fluctuations at the molecular level continuously perturb equilibrium through entropic order, spontaneously deviating from the equilibrium state. In competition energy minimization restores balance (or equilibrium) through entropy disorder. This continuous duel drives all natural processes near equilibrium and forms the basis of the second law of thermodynamics. Fluctuations hold the key in understanding the fundamental mechanisms that achieve equilibrium collectively specially in biological processes (*1*). The key to observing fluctuations lies in the order of fluctuations $\sim 1/\sqrt{N}$ which depends on the population size or degrees of freedom $N$ of the system. These are discernable only for small $N$ at the molecular scales but gets obscured at macroscales. Nanomechanical tools are proving excellent in observing such fluctuational details in biologically relevant problems (*2*–*6*). Fluctuations get mediated by non-equilibrium (NE) energy exchanges with the environment leading from one conformational sub-state to another (*7*–*10*). These appear through entropy order via dissipative pathways and structures in biological systems (*11*, *12*) and is also the basis of biological self-replication (*13*). However, very recently, it has been argued that entropy production does not initiate a process. Instead, dissipative pathways are the result of generalized friction that gets minimized insofar as the constraints allow (*12*), true even along a single stochastic trajectory (*14*). Local force gradients define the constraints of evolution of such pathways. Yet they remain elusive experimentally since at macroscales long range viscous forces from large $N$ smoothens out the transient fluctuational details in the experimentally measurable quantities. But understanding fluctuations is vital to comprehend the complexity inherent in biological processes. We take the example of DNA unravelling. At the theoretical level the features of the rugged energy landscape through which a DNA unravels provides a unifying language for complex bio-mechanisms. The underlying physics is in a similar intriguing entropy duel in the bound and unbound base-pairs (*bp*) of the intermediate bubbles leading to an entropy-driven conformational transition (*15*–*27*). In this work, we re-explore this landscape exploiting nanomechanics with picoliter samples confined in a resonant microcantilver as a tool to understand the fluctuation energetics.

In a DNA the double strands are held together by hydrogen bonds in paired A-T and G-C nucleotides in a sequence. With temperature, the hydrogen bonds break and the strands separate base by base resulting in a complex energy pathway. A bound *bp* is energetically favorable while an unbound state is entropy favored. An intriguing energy competition thus ensues that tries to minimize energy through statistical excursions in a conformation which must be simultaneously entropy favored. The entropy gained in separating the strands must balance the energy cost of breaking base-pairs (*bps*). A complexity in this balance arises from dissimilar energy or entropy contributions of bound and unbound sites for different sequence or pairing combinations (*21–23*). This is because the stability of every *bp* depends on the identity and the orientation of its nearest neighbor *bps* and the number of hydrogen bonds. Cumulatively, they contribute to the thermal instability and complexity in DNA unravelling. Longer sequences add more complexity through cooperativity effects of single stranded regions ("bubbles") bounded in between double stranded sections (*15*, *16*, *24*). The strands eventually separate at higher temperature by the gradual unbinding of the rest of the chain, resulting in a continuous uniform melting/transition profile. The mechanism finds support in Poland-Schrega models (*24*) and improvements that incorporate nonlinear effects (*28*, *29*) or consider cooperativity and nucleation of intermediate bubbles(*15*, *16*, *20*, *30*) through statistical weights. The other principal mechanism argues a rather simpler linear transition, where the strands unpair and unstack like a zipper from one or either end (*25–27*). Essentially, in all cases, the unbinding of a *bp* is treated as a discrete two-state process, irrespective of sequence. Fundamentally, a complex energy-entropy interplay unique to every *bp* unbinding establishes the underlying transitional character in a DNA melting profile.

The entropy-energy contributions in melting transitions are commonly interpreted and analyzed on the premises of equilibrium thermodynamics. This is not because of a matter of choice but since technically the available methods cannot discern the transition dynamics of DNA separately from the dynamics of its surroundings e.g., the heat bath that constitutes the DNA solution. Nevertheless, transition profiles characterised from light absorption at 260-290nm (UV) wavelengths (*17*, *31*), circular dichroism spectroscopy and fluorescence(*32*, *33*), calorimetry (*18*, *19*) or electrophoretic mobility assays (*34*, *35*) provides with a high degree of precision. A Gaussian probability distribution of *bp* dissociation enthalpies readily approximates the equilibrium state. The average change in enthalpy $\langle \Delta H \rangle$ so produced per *bp* unbinding corroborates to an increase in the total average entropy $\langle \Delta S \rangle$ of the system plus the environment. Fundamental relevance of enforcing a normal distribution in the analysis lies with the number of degrees of freedom (DOF) contributing to the energy state levels. The levels constitute the energy path/landscape through which the transition proceeds. Under equilibrium considerations each of the energy levels correspond to an equivalent non-zero statistical ensemble average entropy contribution $\langle \Delta S \rangle = \langle \Delta H \rangle / T \neq 0$, with negligible fluctuations or uncertainties $\left[ \langle \Delta S^2 \rangle - \langle \Delta S \rangle^2 \approx 0 \right]$, $T$ being the temperature of the heat bath. This is especially true for long sequences and high enough concentrations where the DOF contributing to the energy landscape is large (DOF $\rightarrow \infty$) making the distribution uniform.

Reconsider this. A DNA strand ($< 100\,\text{mers}$) melts at temperatures in the range of $320 - 400K$ ($45 - 75^0 C$). This is much lower than the typical average $\langle \Delta H \rangle$ of about 6 *kcal/mol* ($3000K$ molar equivalent) for a strand in the "helix" state (*22–24*, *36*, *37*). Fundamentally, from statistical mechanics considerations, as the strand separates many $\langle s \rangle \sim 10$ DOF are released (different for each A-T and G-C pair) that accounts for a dynamic entropy change per site ($s$ expressed in units of Boltzmann's constant in molar terms *J/mol-K*)(*24, 36*). Result: a net increase in the internal energy. Interestingly, the entropy contribution $-sT$ *J/mol* at room temperature $T \cong 300K$ closely compensates for the loss of binding energy, i.e. $\varepsilon_b \approx -sT$ in the immediate vicinity of the melting temperatures. *But, a base-pair at the boundary between a helix and a melted bubble is neither completely bound nor completely separated. So, a certain decrease in the binding energy is only partially compensated by an incomplete entropy release*. Unavoidably, a dynamic entropy fluctuation scale results at the boundaries, with only few DOF $(< 10)$ contributing to the energy landscape. A fluctuation driven non-equilibrium (NE) consideration thus becomes pertinent. The present experimental study tries to elucidate this. Incidentally, RNA hairpin folding transitions evolving from NE trajectories have been studied by mechanically stretching a single molecule of RNA reversibly and irreversibly between two conformations (*5*, *6*).

NE thermodynamics in general require the description of dynamics of the surroundings (e.g., a heat bath) independent of the system undergoing transition, yet not necessarily enforcing a clear physical system-bath boundary. The key lies in the timescales at which the microscopic variables of the bath become relevant in describing the dynamics (*38–40*). Physically, it is possible to examine this via a measurement system that can follow the energy exchanges with the bath dynamically. Essentially, in the case of DNA unravelling, the energy exchanges would correspond to the entropy release from *bps* on gradual unbinding, and appear as energy dissipated. A microfluidic pL(~100 picolitres) channel cantilever qualifies as a suitable platform to track those changes. Its very low thermal mass presents sensitivity in the order of $mJ/g \cdot K$ corresponding to $\sim pJ$ at $300K$ (*41*), which is crucial to capture the dynamic non-equilibrium binding-unbinding energies $\sim \mu J$ (at $nM$ concentrations) in the melting transition of a DNA strand (Fig 1). The relevance is two-fold. Entropy fluctuations are inherent fast time-scale events that enforce transient relaxations to non-equilibrium states. In competition, a slower energy minimization path smoothens out the fluctuations to achieve a final equilibrium state. The multi-timescale events remain decorrelated by the inherent irreversibility introduced by the fluctuations. Micro-resonator dynamics at resonance provide access to both the timescale events. The faster timescale events are likely to distribute the energy among the DOF $\langle s \rangle$. These statistically manifest as dissipation broadening at resonance. The resonance timescale (slower) on the other hand faithfully captures the evolution of the equilibrium state on complete melting. The importance of the dynamic dissipation and competing energy scales in mechanical resonators has been raised (*42*). Earlier discourses (*39*, *43*) extendedly elucidated this in the context of information/entropy change central to a NE transition. The competition becomes apparent in a multi-stage transition profile evident from the experimental results (Fig 1). The competing scales signify partition function contributions from an evolving fraction of open-bound *bps* of DNA strands and the contribution of completely separated strands to the energy landscape. We find it convenient to approximate this in terms of a two-stage Boltzmann distribution where the average contribution from open links relate to the relative change in dynamic dissipation $D$ (Fig 1) as

$$\frac{D}{D_{\min}} = Z\mathrm{f}\left(\frac{\exp(-\beta\Delta_1)}{1-\exp(-\beta\Delta_1)}\right)^{\mathrm{f}} + (1-Z)\frac{\exp(-\beta\Delta_2)}{1-\exp(-\beta\Delta_2)}\,. \tag{1}$$

Here **f** is the fractional contribution of individual DNA strands, $\beta\left(=1/k_B T\right)$ is the energy inverse Lagrange multiplier, $\Delta_1$ and $\Delta_2$ are the dynamic non-equilibrium energy differences of the two stages of transition and $Z$ is the span of measured dissipation landscape. Essentially, **f** becomes a measure of the fluctuations in entropy $\langle s \rangle$ released at the *bp* unbinding boundaries.

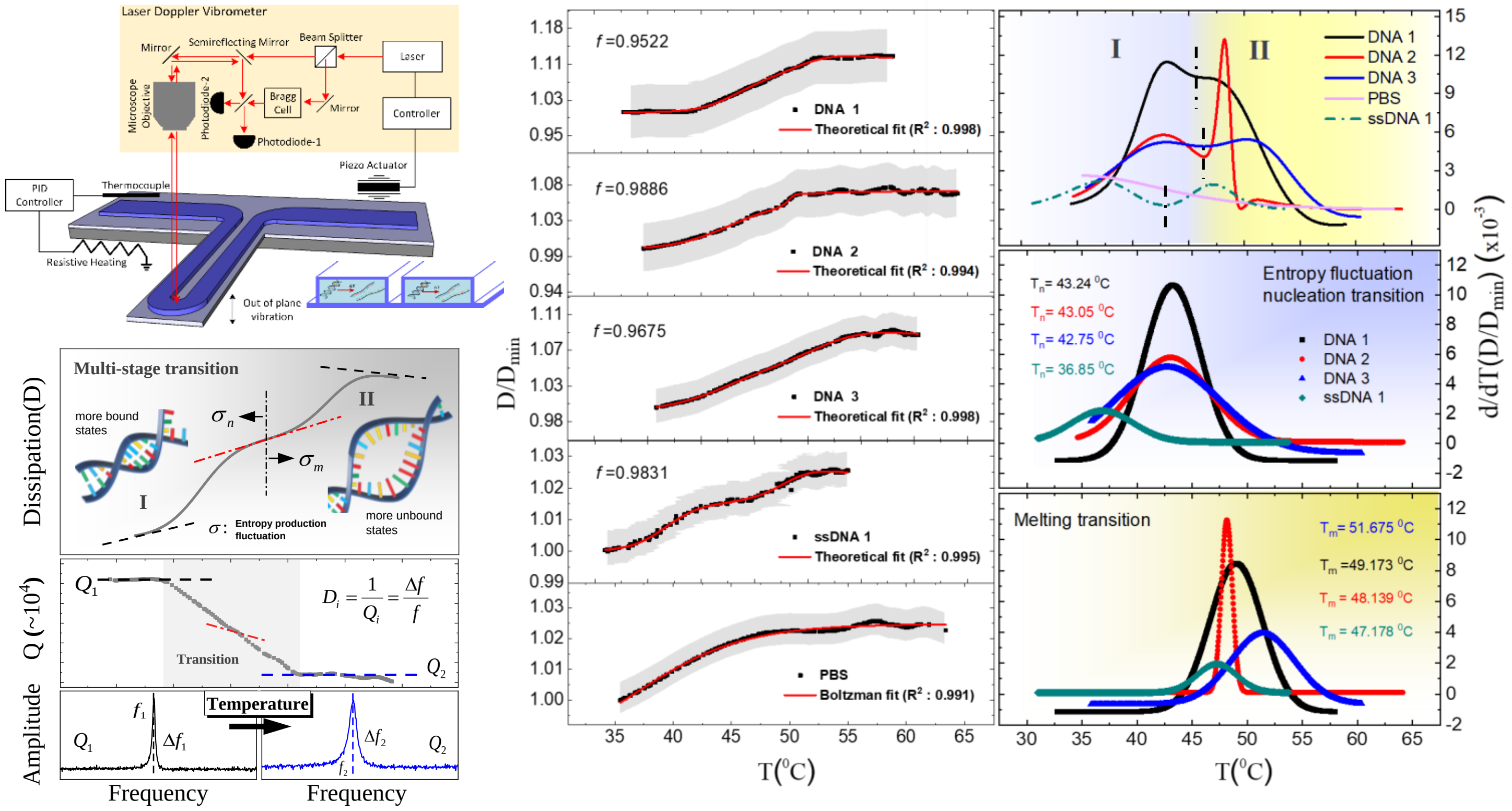

**Fig. 1.** Experimental schematic and results of dynamic dissipation changes of a microfluidic channel cantilever at resonance for DNA solutions of varying oligomer lengths and base-pair moieties. Theoretical fits using Eq 1 and the derivatives of the fitted profiles reveal an apparent two-stage transition indicated by I and II regions in the graphs. Fitting parameter f for respective DNA oligomers is also indicated. The gray bands represent the standard error. The transitions are demodulated with Gaussian distributions and corresponding obtained transition temperatures are shown. The emergence of a fluctuation scale at pre-melting temperatures is evident and the respective possibilities of ordering and disordering states of the DNA strands from fluctuations are denoted in the schematic.

| | |
|---|---|
| **DNA-1** | 5'-AAAAAAATTTTTAAATTTAAATTTAAATTTAAATTTAAATTTAAA-3' |
| **DNA-2** | 5'-CCCAAATTCAAATTAATAAAAACCC-3' |
| **DNA-3** | 5'-GGAGGGAAGAAGAAATTGGGGTTGA-3' |

The competition originating from fluctuations apparently drive the transition to either a nucleation - entropy order at $T_n$ or to melting - entropy disorder at $T_m$ (Fig 1). The entropy order transition has more significance from a NE view-point. The energy cost from loosing DOF in ordering/binding of *bps* is compensated by dissipating energy (exchanging heat with the bath irreversibly) in transient relaxations to NE states. Energy associated with each DOF are statistically independent. Statistical excursions thus result in an energy relaxation timescale $\tau_i$ over which an entropy change of $\langle \sigma_i \rangle = \langle s \rangle \cdot \tau_i$ is affected. This is validated by the simultaneous measurement of dynamic steady state temperature inflections at definite timescales $\tau$ respectively for each DNA solution and the PBS medium (Fig 2). Incoherence drives the relaxation to proceed through a distribution. Gaussian broadening ($\Gamma_n$ and $\Gamma_m$) approximation is enforced to de-convolute the resulting entropy production $\sigma_n$ and $\sigma_m$ distributions in either case (Fig 1). This is done under the assumption that many statistical events result in entropy $\langle s \rangle$ fluctuations constituting a partial release.

Essentially, the transition width signifies the interplay of DOF providing a measure of the degree of temporal order in the evolving dynamics. From the transition characteristics (Fig 1) we may define a dynamic transitional-ordering parameter $\lambda = (f - A_n)/(1 - A_m)$ as an ordering predictability scale where $A_n$ and $A_m$ are the areas under nucleation/ordering and melting/disordering transition peaks respectively. The significance of $\lambda$ can be appreciated from the understanding of Kolmogorov-Sinai (KS) entropy rates that originate from finite time trajectories (*38*, *44*). The energy distributed among each DOF in binding/unbinding can in general be associated to positive Lyapunov exponents (*38*, *44–46*) that are rather difficult to compute. Resonance broadening effectively captures the net effect unique to the interplay among the DOF. It can thus be a measure of a predictable scale of the dynamics. The experimental results give $0.92 < \lambda < 0.96$ for the three different DNA sequences and the ssDNA-1 oligomer. Interestingly, invoking a relation between the predictability scale $\lambda$ and physical entropy $\langle s \rangle$ using (*44*) gives $[\lambda] \equiv$ a dynamic entropy range $7 < \langle s \rangle < 8$. This confirms hypothesis of a fluctuation scale $\sim 70-80\%$ contributing to the observed ordering transitions, the fluctuations corresponding to partial entropy release $\sim \text{DOF} < 10$. Consequently, in the limit $\lambda \to 1$, $\langle s \rangle \to 9$ using (*44*), suggesting a fluctuation scale of $90\%$. Thus, we conclude that fluctuations favor a dynamic entropy order at $T_n$ in the energy landscapes of the DNA transitions. This is in line with the conclusions of higher KS entropy from dynamic ordering (*38*, *47*).

Our conclusions find justification in the observance of a higher entropy order probability $P_{\sigma_n} = \Gamma_n / T_n$ over entropy disorder probability $P_{\sigma_m} = \Gamma_m / T_m$ as a function of higher $\lambda$, evident from the exponential dependence shown in Fig 3(a). $P_\sigma$ 's are expected to satisfy the generalized symmetry relation known by fluctuation theorems(*45*, *46*, *48*, *49*)

$$\frac{P_{\sigma_n}}{P_{\sigma_m}} = \exp\left(\frac{\langle \sigma \rangle}{k_B \tau}\right) \tag{2}$$

where $\sigma_n \approx \sigma_m \approx \sigma$ are of the same order and $k_B$ is Boltzmann's constant. The transition probabilities reveal the outcome of the intriguing competition as shown in Fig 3(a). The exponential nature establishes a correspondence of the time dependent entropy production to the dynamic ordering-parameter $\lambda$ suggesting that the ordering becomes a higher predictable state compared to the disordering state at $T_n < T_m$.

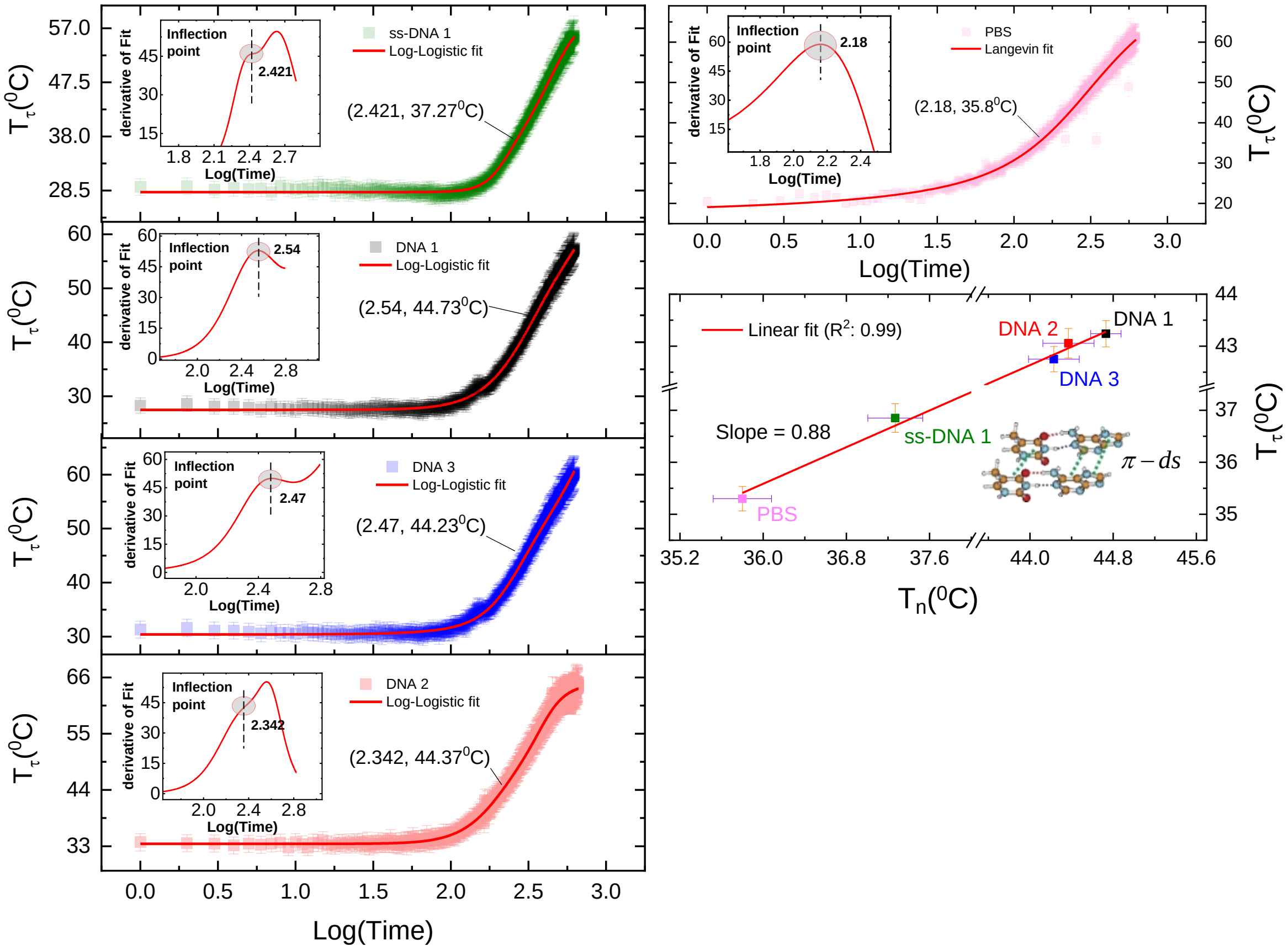

**Fig 2.** Dynamic temperature vs time evolution with corresponding error bars indicating temperature inflections of the bath at definite time-scales. The temperature inflections correspond to the ordering transition temperatures linearly as shown. Ordering in ss-DNA 1 correspond to $\pi -$stacking as indicated.

A direct conclusion: the possibility of an alternate pathway to DNA melting transition. The energy cost of complete unbinding of strands is high. This can be compensated by the energy released from higher entropy order in nucleation. The denatured bubbles may thus collapse forming an intermediate center-bound zipper as shown in Fig 3(b). The energy released from entropy ordering gets utilized in unbinding the ends of the zipper which incidentally is an energy demanding step. Further insight to the above argument can be drawn from transition width of DNA 2 oligomer (Fig 1) that has G-C pairs on both ends of the chain. Termination in G-C *bps* would make DNA 2 oligomer highest in energy demand for complete unbinding compared to the other oligomer chains with at least one A-T termination. Result: sharpest melting transition (Fig 1). In general, the onset of an entropy fluctuation scale near the transition suggests the overlaying of a secondary transition mechanism over a cooperative first order transition(1–20) through entropy order. This has to achieve through a relative entropy production maximization(*50–53*), compensating for the partial losses at the boundaries of dissociating *bps* (micro-canonical regime). This is possible only where fluctuations dominate. This hypothesis is evidently justified in the relative interplay of time dependant ordering and disordering probabilities $P_{\sigma_n}/P_{\sigma_m}$ as a function of concentration and its derivate plot (Fig 4).

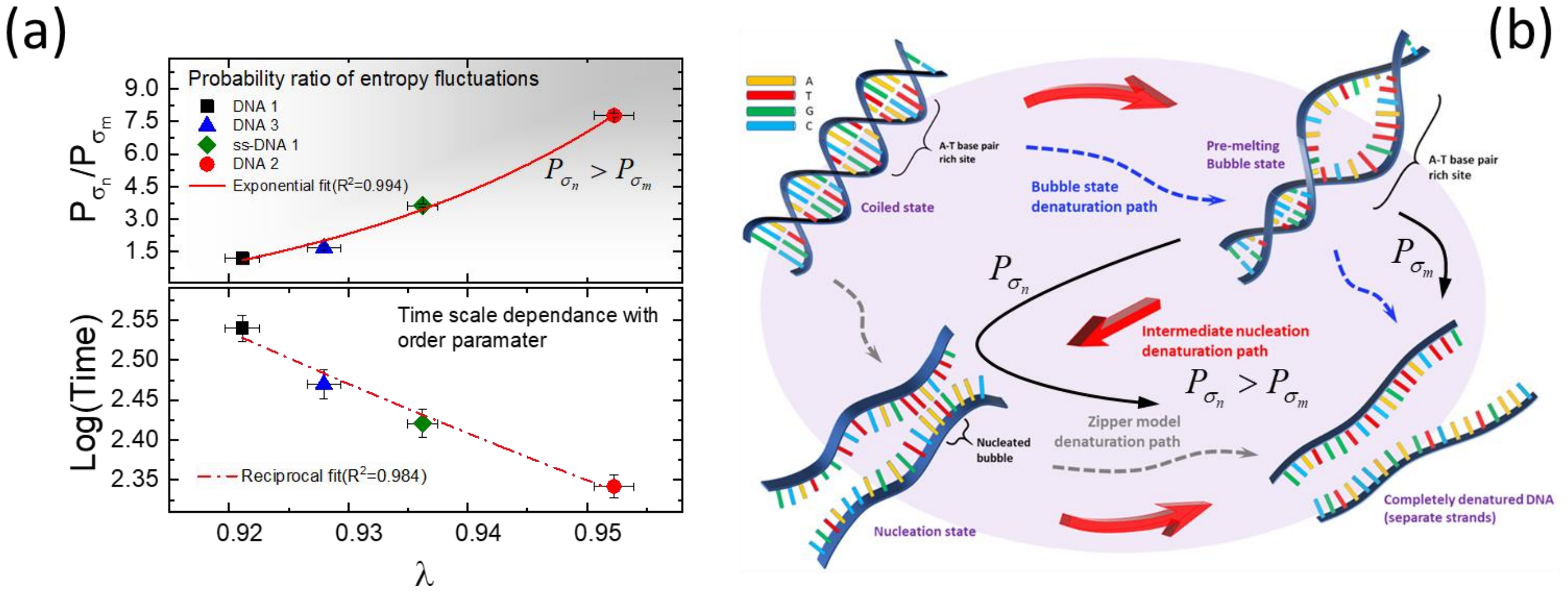

**Fig. 3.** Probability rates and transition pathway schematic. (a) Relative entropy probability rates and (b) suggested alternate transition pathway justifying the higher ordering entropy production relative to the disordering entropy production.

Higher concentrations would result in a crossover into the thermodynamic limit where the fluctuation scale would cease to compete over the ensemble average scale (*48*). This is apparent in an exponential decline of the entropy production with concentration (Fig 4), justifying the usual normal distribution of energy states in the thermodynamic limit. The dynamic change in entropy in NE transitions is expected to manifest through a measure of the difference in energy exchanged with the system irreversibly $\propto \Delta T = T_m - T_n$. This is achievable at the cost of net transition broadening $\Gamma_n + \Gamma_m$ through probability excursions of decorrelated entropy events. A relative entropy fluctuation interplay maximization $\Sigma_{tr} = \Delta T/(\Gamma_n + \Gamma_m)$ thus evolves as shown in Fig 4 (* denotes a normalized scale). Eventual achievement of a steady state in the disordered melted strands gets reflected in the relative dissipation change $\Delta D_{eq}$. This is expected to be proportional to the steady state entropy change (*49*).

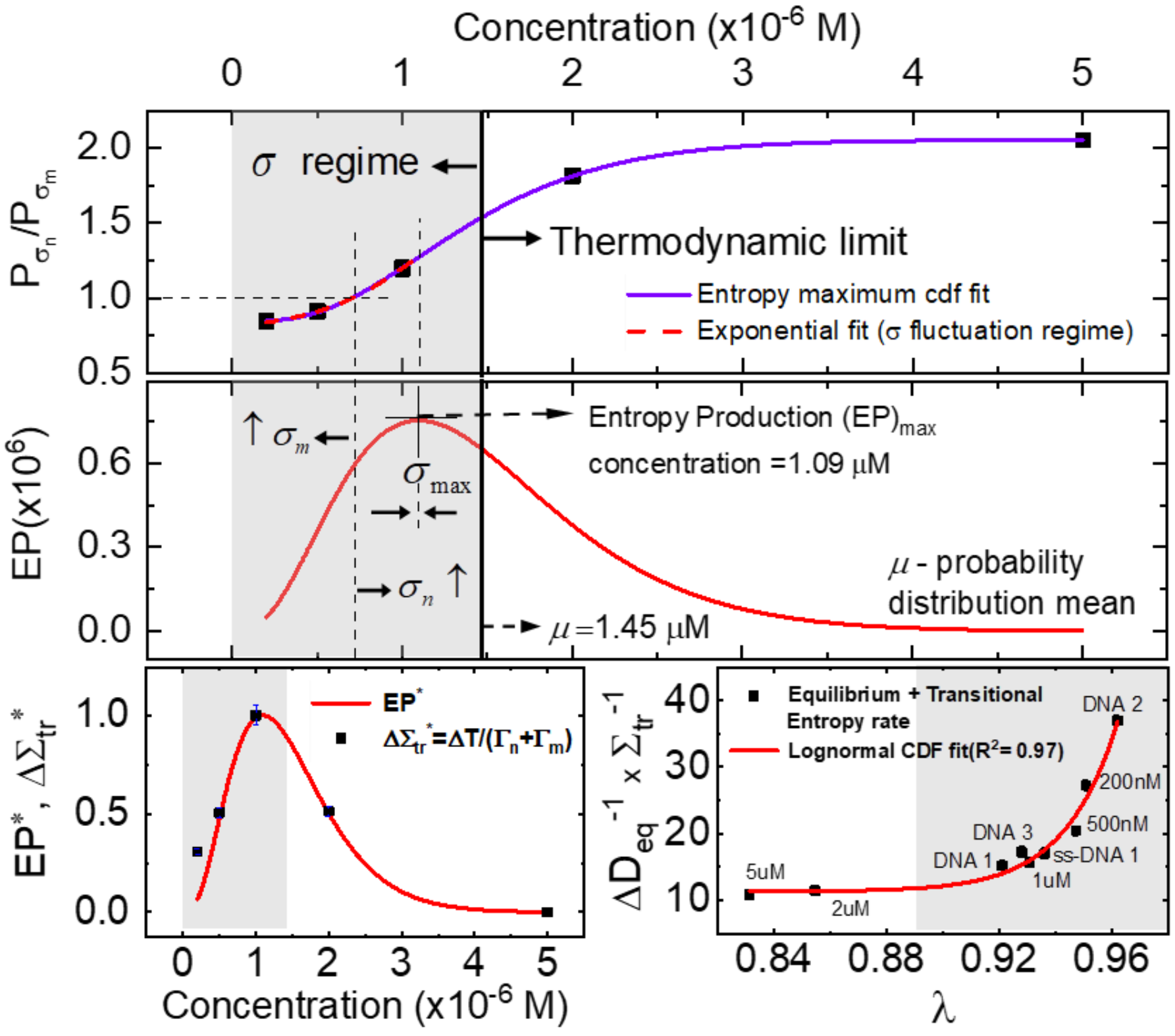

**Fig. 4.** Entropy rate and concentration dependence. Entropy maximization naturally selects a NE transition pathway with the net entropy rate varying exponentially with the ordering parameter $\lambda$.

Intrinsically, the entropy changes in NE transition states and that in the equilibrium steady states are time decorrelated. Thus, the net entropy production rate $1/(\Sigma_{tr} + \Delta D_{eq})$ would be the product $(\Sigma_{tr} \cdot \Delta D_{eq})^{-1}$ (*54*), representing a cumulative density function of entropy production. Gibbsian hypothesis predicts that the only function that can satisfy the above equality is the exponential in the form $(\Sigma_{tr} \cdot \Delta D_{eq})^{-1} \propto \exp(\lambda)$, where $\lambda$ would represent a dynamic measure that is independent of the either energy scale considerations on the left hand side of the above equation. The experimental outcome perfectly attests to this fundamental postulate as clear from Fig 4 with $\lambda$ satisfying the condition of a dynamic fluctuation ordering-parameter defining the transitional characteristics. In other words, a dynamic fluctuation feedback scale evolves in DNA melting transition that tries to enforce dynamic equilibrium states through KS entropy ordering maximization over disordering (*50–53*). Disordering takes over in the thermodynamic limit or when the microscopic variables of the heat bath dictate the evolving dynamics making $\lambda$ small (Fig 4). Essentially, the fluctuations enforce dynamic instability displacing the system from equilibrium. Entropy maximization feedback enforces the system to evolve under NE conditions until a final steady state is reached (*47*). The transitional character of melting profile from our experiments reveal linked fluctuation energy standards associated with the stability of a DNA strand. Studying the dynamic NE transition through cantilever dynamics thus opens up a novel way to understand the cooperativity effects of near neighbor *bps* in DNA transition, specially in the pre-melting regime ($< 42^0 C$) which is biologically relevant. Finer aspects of fluctuation driven entropy ordering, and the suggested intermediate pathway similar to (*55*) may shed light on the physics of gene propagation through DNA transcription and replication.

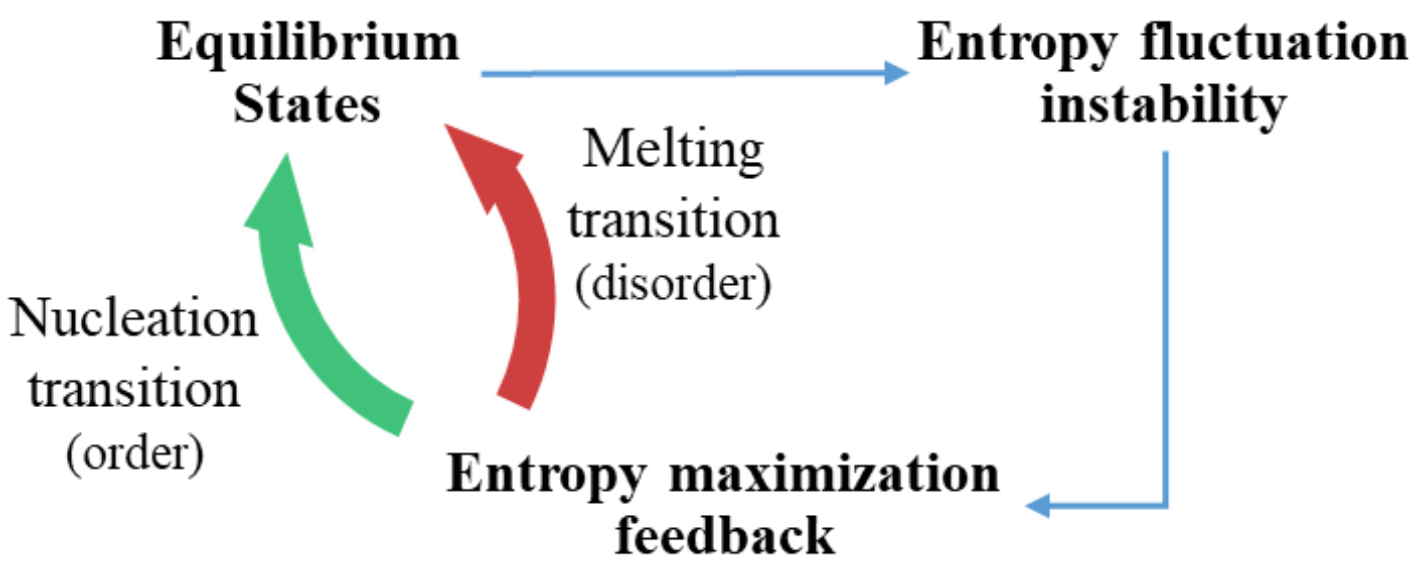

**Acknowledgements:**

**Funding:** This research has been funded by Canada Excellence Research Chair (CERC) for Oil-Sands Molecular Engineering at University of Alberta. T.T. and A.P. were partially funded by RENEW Institute, University at Buffalo and School of Engineering and Applied Sciences, University at Buffalo.

**Author Contributions:** K.J. synthesized the dsDNA samples for experiments, K.J. and F.K. designed and performed experiments with assistance from J.T. The research was supervised by T.T. A.P. and T.T were involved in study design. A.P. analysed the data, performed the theoretical formulation and simulations, and prepared graphs and display items. A.P. and T.T. wrote the manuscript. All authors discussed results and commented on the manuscript.

**Competing interests:** The authors declare no conflict of interest.

**Fig. 1.** Experimental schematic and results of dynamic dissipation changes of a microfluidic channel cantilever at resonance for DNA solutions of varying oligomer lengths and base-pair moieties. Theoretical fits using Eq 1 and the derivatives of the fitted profiles reveal an apparent two-stage transition indicated by I and II regions in the graphs. Fitting parameter f for respective DNA oligomers is also indicated. The gray bands represent the standard error. The transitions are demodulated with Gaussian distributions and corresponding obtained transition temperatures are shown. The emergence of a fluctuation scale at pre-melting temperatures is evident and the respective possibilities of ordering and disordering states of the DNA strands from fluctuations are denoted in the schematic.

| | |
|---|---|
| **DNA-1** | 5'-AAAAAAATTTTTAAATTTAAATTTAAATTTAAATTTAAATTTAAA-3' |
| **DNA-2** | 5'-CCCAAATTCAAATTAATAAAAACCC-3' |
| **DNA-3** | 5'-GGAGGGAAGAAGAAATTGGGGTTGA-3' |

**Fig. 2.** Dynamic temperature vs time evolution with corresponding error bars indicating temperature inflections of the bath at definite time-scales. The temperature inflections correspond to the ordering transition temperatures linearly as shown. Ordering in ss-DNA 1 correspond to $\pi$-stacking as indicated.

**Fig. 3.** Probability rates and transition pathway schematic. (a) Relative entropy probability rates and (b) suggested alternate transition pathway justifying the higher ordering entropy production relative to the disordering entropy production.

**Fig. 4.** Entropy rate and concentration dependence. Entropy maximization naturally selects a NE transition pathway with the net entropy rate varying exponentially with the ordering parameter $\lambda$.

**Materials and Methods:** (for online and HTML presentation only; detailed methods provided in Supplementary Information)

**dsDNA samples preparation**

All the DNA samples were purchased from Integrated DNA Technologies (IDT, Coralville, IA, USA). Freeze-dried ssDNA samples were first dissolved in calculated amounts of phosphate buffer saline (PBS, Sigma-Aldrich, Oakville, ON, Canada) as $100\,\mu M$ stock solutions before further dilution. dsDNA samples were prepared by incubating the ssDNA's and their complimentary strands at room temperature with respective calculated concentrations.

**Fabrication, Packaging and Experimental technique**

The microfluidic cantilever (MC516, Fourien Inc. AB, Canada) was fabricated using silicon nitride (SiN) as a structural material. To get consistent properties of deposition of the thin film of SiN, the deposition was performed using low pressure chemical vapor deposition (LPCVD). The $500\,\mu m$ long (overhang length) and $57\,\mu m$ wide cantilever incorporates a microfluidic channel, which is $16\,\mu m$ in width and $980\,\mu m$ in overall length. The DNA samples (confined in the channel) were heated using a $25W$, $5\,\Omega$ resistive heater (KAL25F, Stackpole Electronics Inc., Raleigh, NC, USA). Miniature size of a microfluidic cantilever makes it an ideal platform for measuring changes in specific gravities of picolitres (pL) of confined liquids by continuous monitoring of its resonance frequency. In addition to quantifying the mass change (and the density) of the confined liquid sample, resonance response can also be used for monitoring the quality factor (Q) of the cantilever, a unit less parameter which describes the rate at which energy gets dissipated per oscillation. The sensor chip was placed in a vacuum holder and operated at a $10^{-5}\,mTorr$ pressure in order to reduce air damping thus increasing the Q to desired levels for higher sensitivity. The resonance of the DNA filled cantilever was measured by a MSA-500 laser Doppler vibrometer (LDV Polytec, Irvine, CA, USA). The resonance frequency and Q were simultaneously measured. dsDNA sequences with variations in base pair numbers and/or composition (G-C ratio) were analyzed. A more detailed description is provided in the Supplementary Information file available online.

# Supplementary Material

**This Supplementary Material includes:**

Supplementary text

Figs. S1.1 to S6.3

References *(56-67)*

**S1. UV-Vis $OD_{260}$ thermal stability experiments vs Dissipation results:**

$OD_{260}$ measurement was performed using a Varian Cary 50 UV-vis spectrometer (Agilent, Santa Clara, CA, USA) equipped with an Isotemp 3016D heated bath circulator (Fisher Scientific, Ottawa, ON, Canada). Incubated dsDNA samples were tested using a capped 1 mL quartz cuvette. The temperature was ramped according to the estimated $T_m$ . 260 nm absorption peaks were utilized for quantifying the DNA samples concentration variation during the melting process. $T_m$ was determined from the fitted results of Abs (absorption in normalized units) and their subsequent derivatives d(Abs)/dT as shown.

Results in figure depict UV-Vis $OD_{260}$ measurement of the DNA samples from $40$ to $55^0C$. Corresponding transition curve fits and their derivatives reveal the transition maxima for different samples. The $OD_{260}$ responses nicely corroborates to data obtained from dissipation experiments confirming entropy disordering transition hypothesis at higher temperatures. E.g., $T_m$ of dsDNA1 (DNA 1 in Fig S1) is determined as $48.83^0C$ from $OD_{260}$ that corroborates to dynamic dissipation measurement result of $49.17^0C$ with a deviation percentage of 0.7%. The corresponding $T_m$'s of DNA samples of different sequences were also determined by $OD_{260}$ and were correspondingly matched as shown in the bar graph with corresponding error bars. dsDNA 1 shows a sharp absorption increase between $46$ to $51^0C$ that can be accounted from an entropy disordering transition as explained in the main article. Interestingly, ssDNA-1 does not show such a sharp absorption change with temperature ramping. The CDF fit (as shown) however, can explain the possibility of a nonlinear decrease in $\pi-\pi$ overlapping or gradual unbinding of longer self-hybridized chains(*56*, *57*) as a function of increasing bath temperature. The apparent sharp transition of DNA 2 oligomer from $OD_{260}$ matches the transition observed in the dissipation results. In advantage, a sharper transition peak from the dissipation results shows higher resolvability character of dynamic dissipation measurements. Consequently, it makes the fluctuation driven entropy ordering-disordering argument stronger. The alternate transition pathway through intermediate nucleated or folded stages as suggested by our results is similar to(*55*).

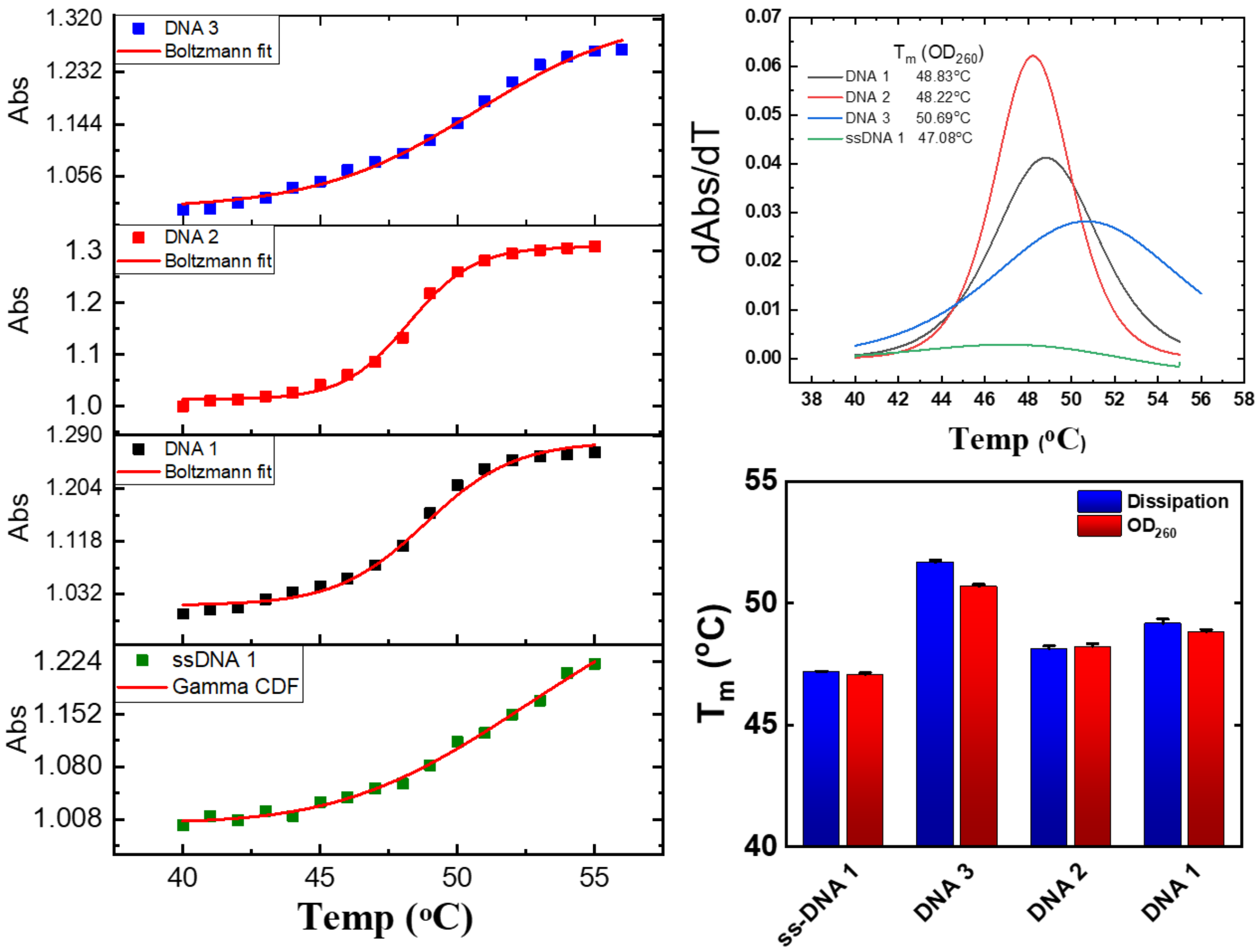

**Fig. S1.1|** Comparison of Melting Transition peak maxima results from Dissipation experiments with $OD_{260}$ thermal stability experimental results (left pane). The bar graphs on bottom right pane compares results. Respective error bars also shown.

**S2. Discussion on possible intermediate nucleation structures in the transition pathway:**

A possible approach to understanding the entropy ordering is the consideration of intermediate folded or nucleation conformations similar to(*55*). The ordering may proceed through collapse of intermediate bubbles to form centre bound zipper as depicted below. Some of the possible conformations for the different oligomers are represented below.

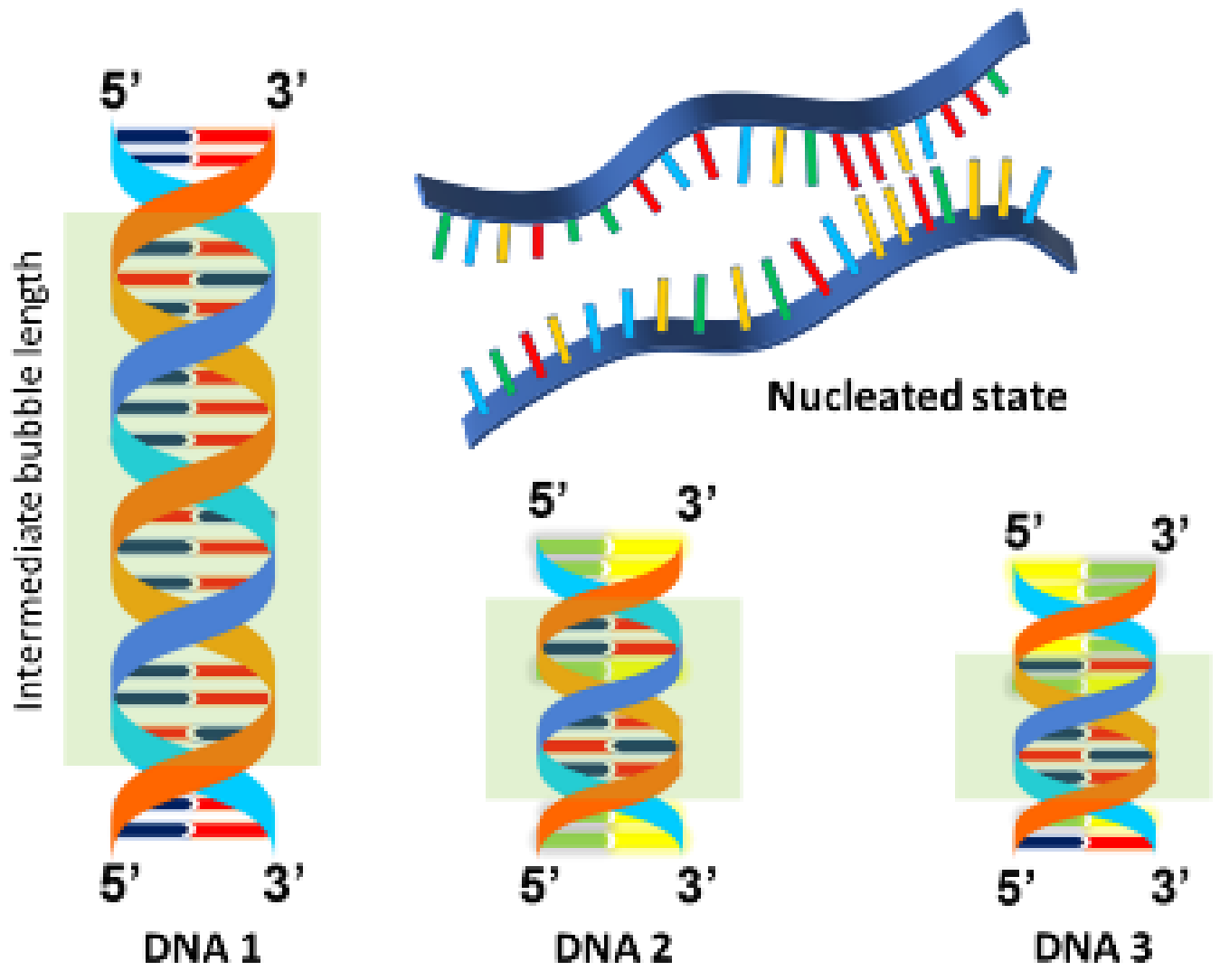

**Fig. S2.1|** Possible bubble regions for the three oligomers and a representative centre bound nucleated zipper from entropy ordering. The green shaded regions represent possible bubble state locations. The bubble lengths play a major role in determining the energy landscape as known(*15*, *16*). Our hypothesis complements the existing knowledge augmenting the possibility of an intermediate nucleated stage forming centre bound zipper as shown. The other possibility is self-hybridized hairpins as shown below.

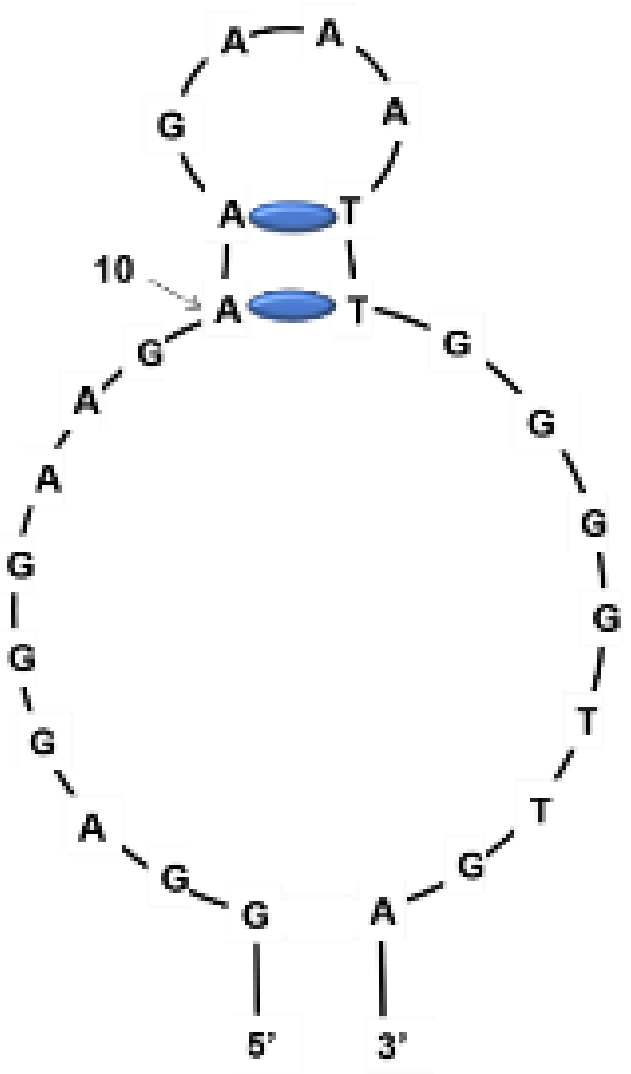

**Fig. S2.2|** A possible self-hybridized hairpin structure of DNA 3. The H-bonds forms between A/T at 10/17 and 11/16, lead to a large decrease in transition maxima from the theoretical estimation $T_m \sim 52^0 C$ for the DNA 3 sequence. The estimated $T_m$ for the above hairpin is $\sim 26^0 C$ making it a very unstable configuration at around room temperature. This is even much lower than the observed nucleation transition maxima $T_n \sim 42^0 C$, a temperature range that is usually considered biologically relevant. Deviation from the estimate for the above hairpin supports the argument of an intermediate nucleation of bubbles to form a centre bound zipper conformation (as shown in Fig S2.1) augmenting an intermediate nucleation stage in the transition pathway as shown in Fig 3b of article, rather than a hairpin structure conformation.

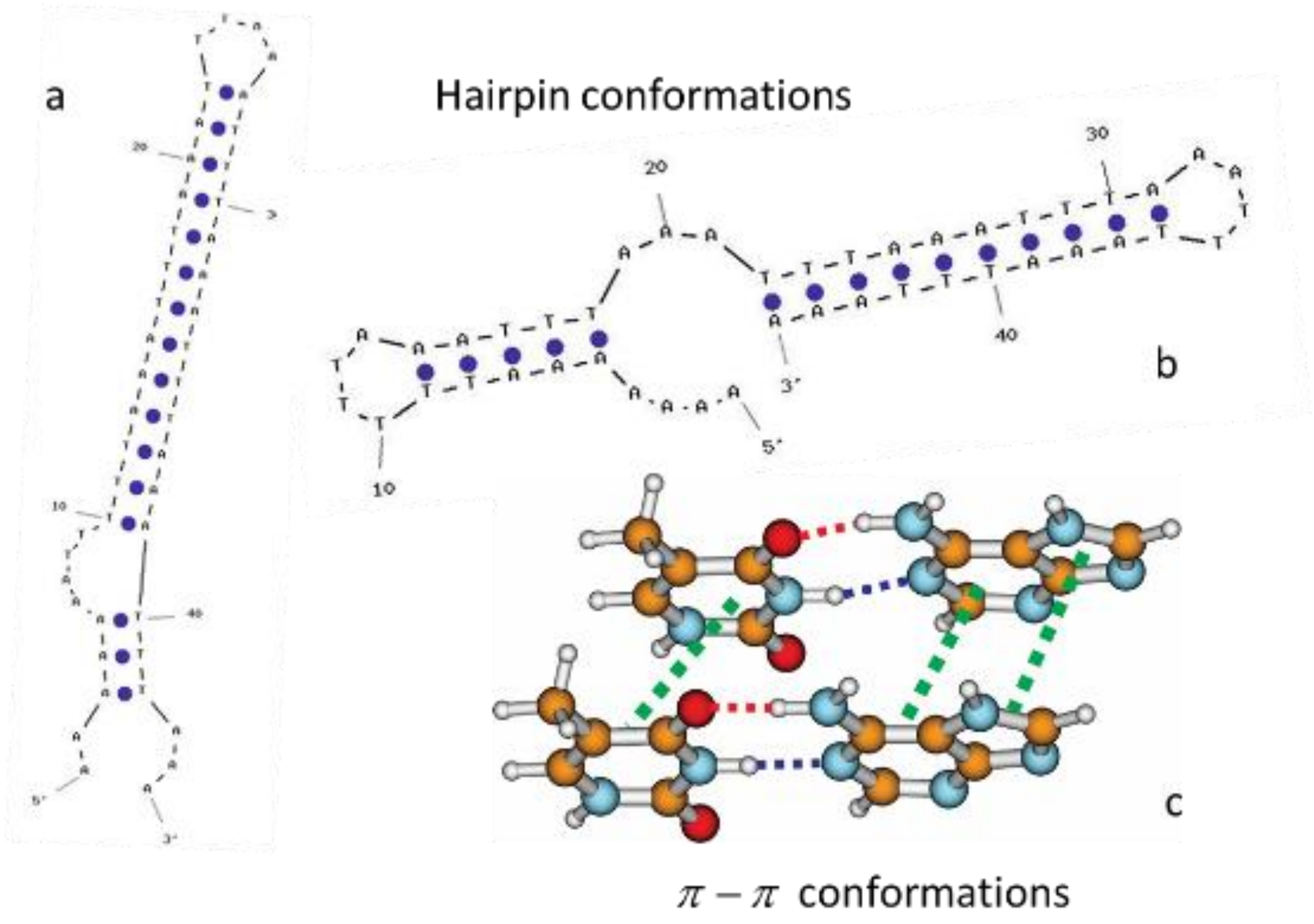

**Fig. S2.3|** Two possible hairpin structures (a) and (b) for ssDNA 1 as determined by OligoAnalyzer 3.1. These may represent the intermediate nucleation transition stage. The estimated melting temperature $T_m$ for the above configurations are in the range of $42-50^0C$. Alternatively, $\pi-\pi$ stacked dimer conformations of ssDNA 1 strands at interaction energy scales in the range of $3.82-5.37\,kcal/mol$ (*37*) is possible, which corresponds to the same range. This readily matches the observed $T_m \sim 47^0C$ in our dissipation and optical density experiments. The observed nucleation transition $T_n \sim 36^0C$ corresponds to an energy scale of $\sim 3\,kcal/mol$ that is equivalent to the average difference in energy scales between paired A-T ($7\,kcal/mol$) and stacked A-T conformations ($3.82-5.37\,kcal/mol$)(*37*). This probably appears from entropy fluctuations in dimer stacking or pairing. Such a transition provides an added confirmation of fluctuation driven entropy ordering in the transition pathway. The melting transition proceeds more like a phase transition from dimer stacked configuration to unstacked monomers in case of ssDNA 1 rather than *bp* unbinding transition as observed for the rest of the oligomers. (*55*)provides added proof of such varied intermediate folded conformations that essentially originates from entropy ordering. The role of ordering as in intermediate stage in the transition pathway thus becomes very apparent.

**S3. Allan deviation, Stability and Error analysis:**

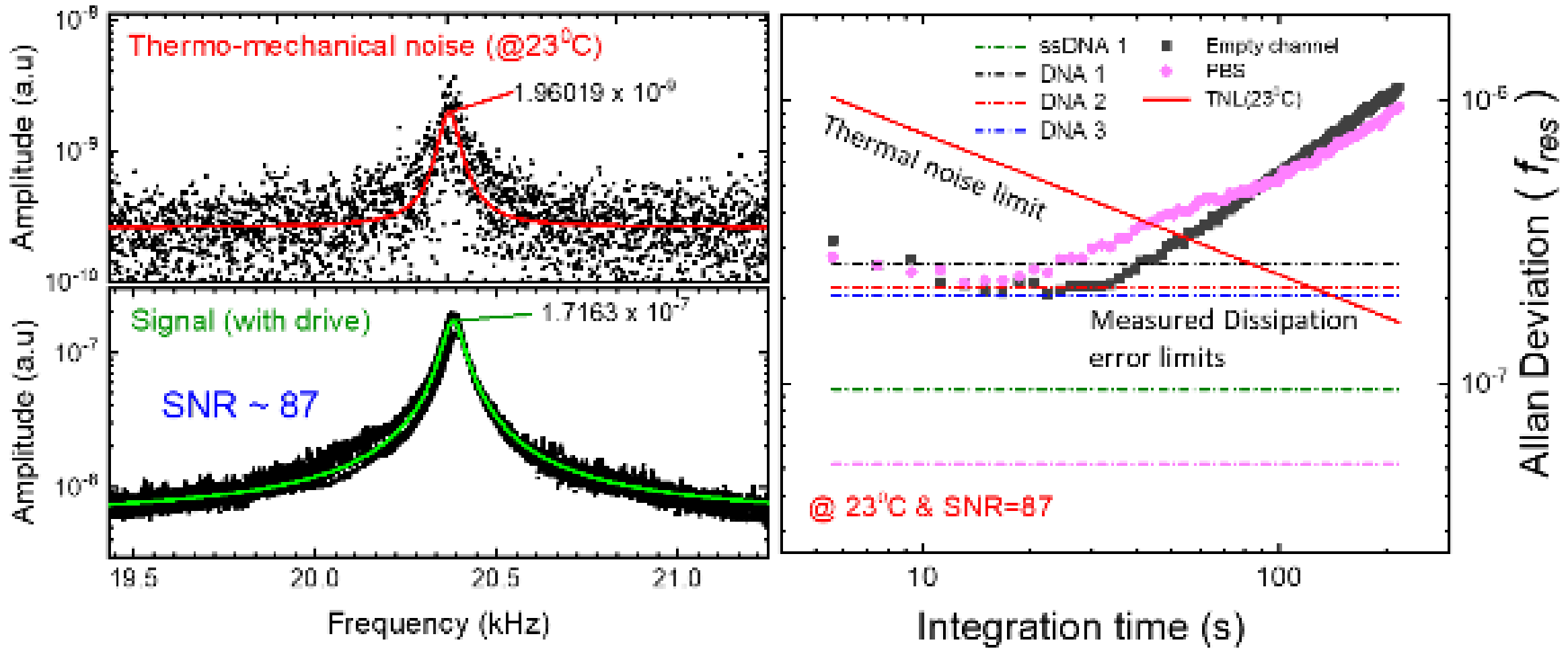

**Fig. S3.1|** Thermo-mechanical noise and Allan deviation at $23^0C$. The solid red line represents the thermal noise limit computed for Q ~ 9000 with an $SNR_{amp}$ $\sim 87$ as determined(*58*). The resonator timescale $\tau_{res} = 2Q/f_{res} \sim 0.8s$. Time sampling scale used in experiments was $1.8s$. This is at par with the general practice of capturing enough sample points that can faithfully reconstruct response corresponding to five time-constants – enough time to provide with a steady state response data. An integration timescale $\sim 4s$ is enough to capture dynamical data with the least error as shown. Systematic deviations start beyond $30s$. The measured error limits for dissipation across different oligomers are plotted as dash-dotted horizontal lines as shown which are well within the same order of magnitude of Allan deviations depicting frequency stability.

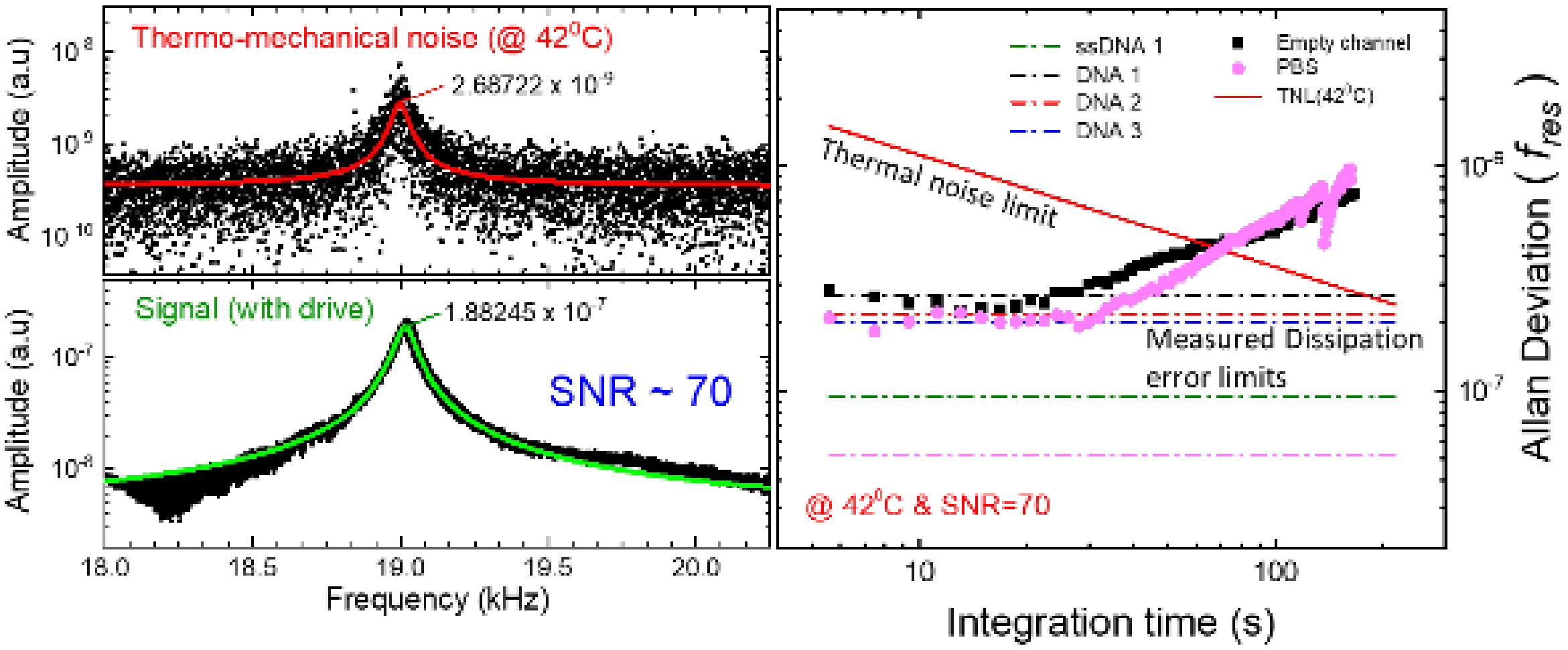

**Fig. S3.2|** Thermo-mechanical noise and Allan deviation at $42^0C$. The solid red line represents the thermal noise limit where Q ~ 8500 and $SNR_{amp}$ ~ 70 as determined.

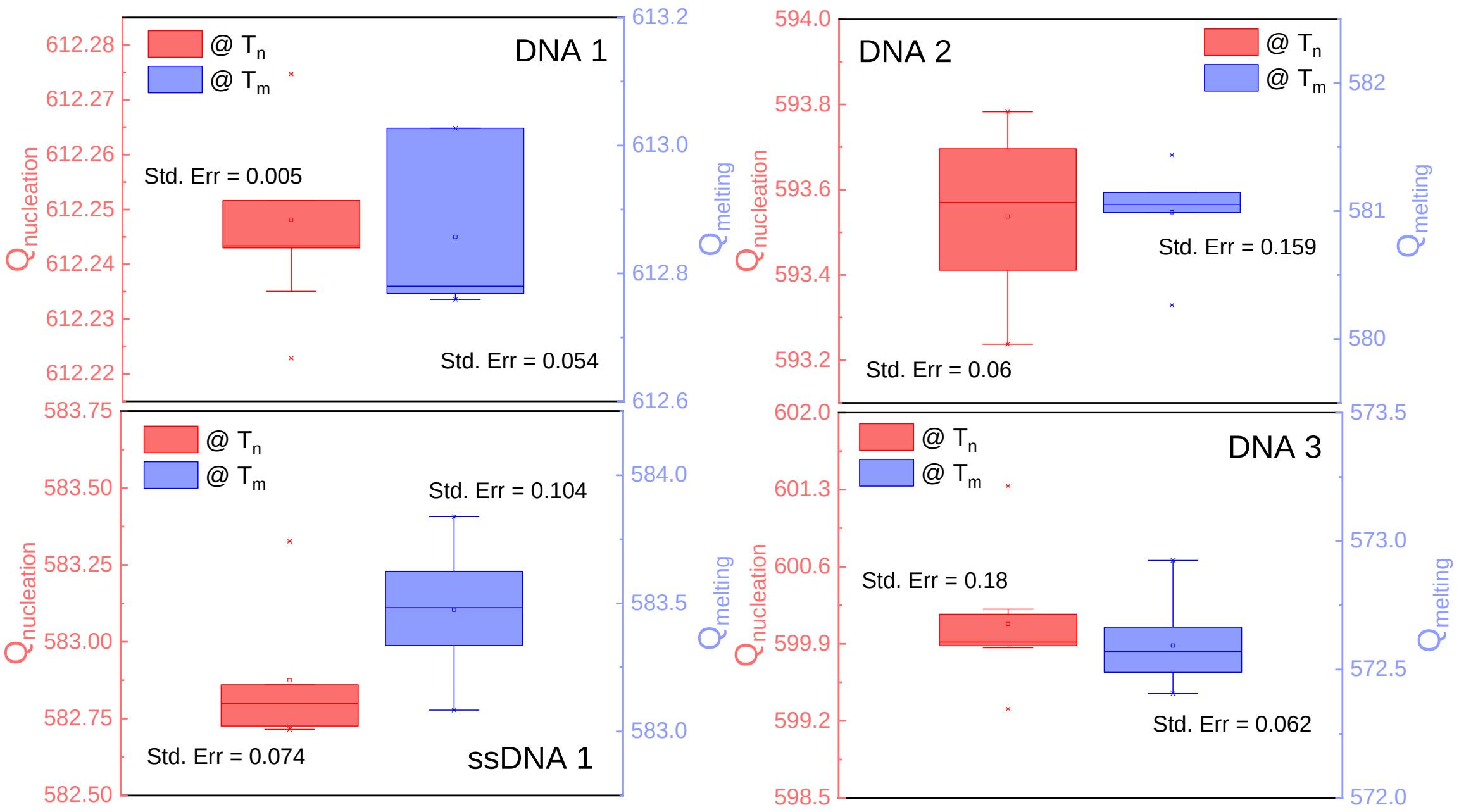

**Fig. S3.3|** Statistical estimation of standard error of measured Q at nucleation and melting transition maxima for the different oligomers. The error percentages in the measurements are well below the estimated thermal noise limits as shown in Fig S3.2.

## S4. DNA 1 concentration analysis and transition stage demodulations:

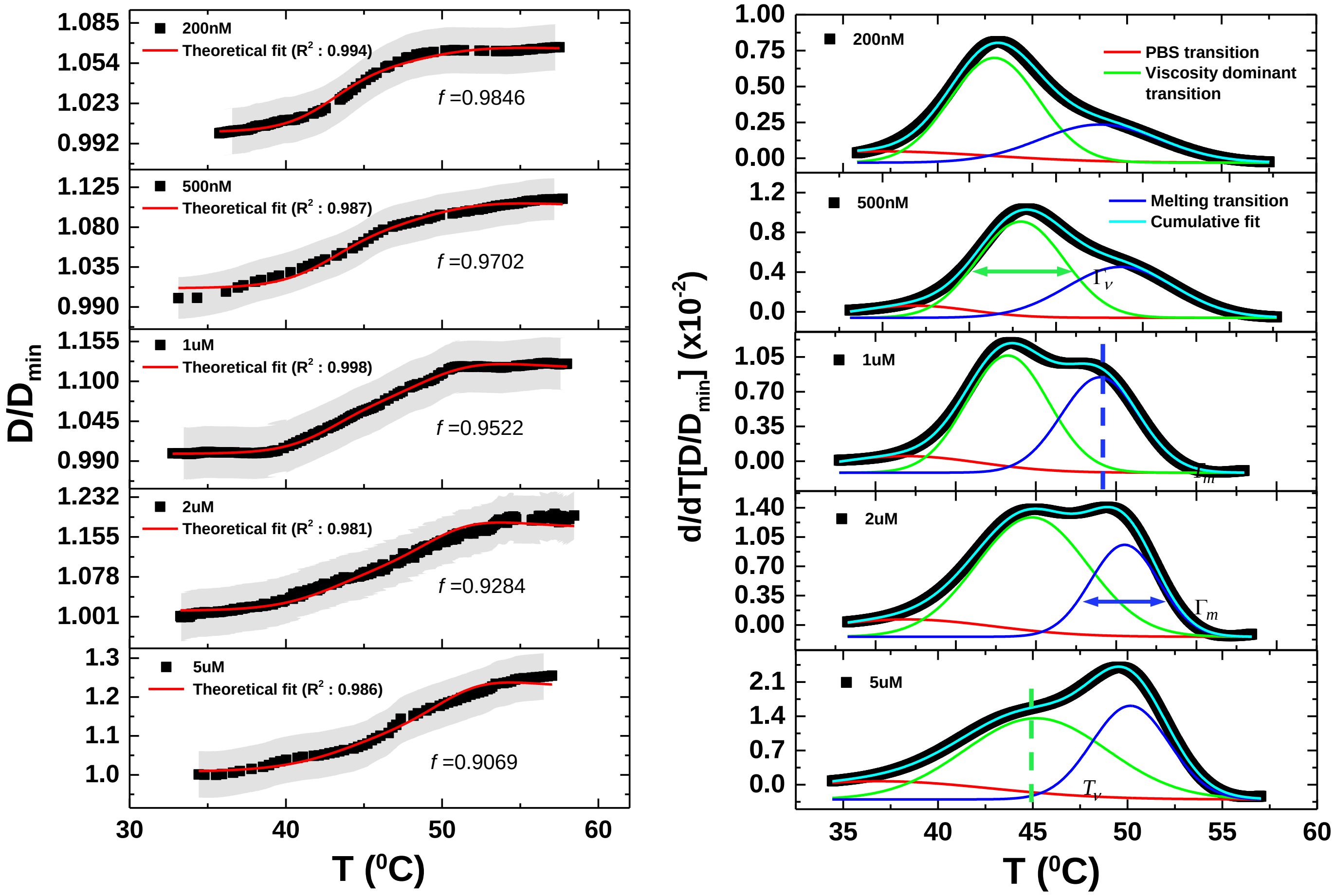

**Fig. S4.1|** Dissipation change as a function of temperature with corresponding theoretical fits. The right pane shows the demodulated transition peaks from the derivative plots of the theoretical fits. The relative competition of ordering/nucleation and disorder/melting transitions as a function of concentration is apparent.

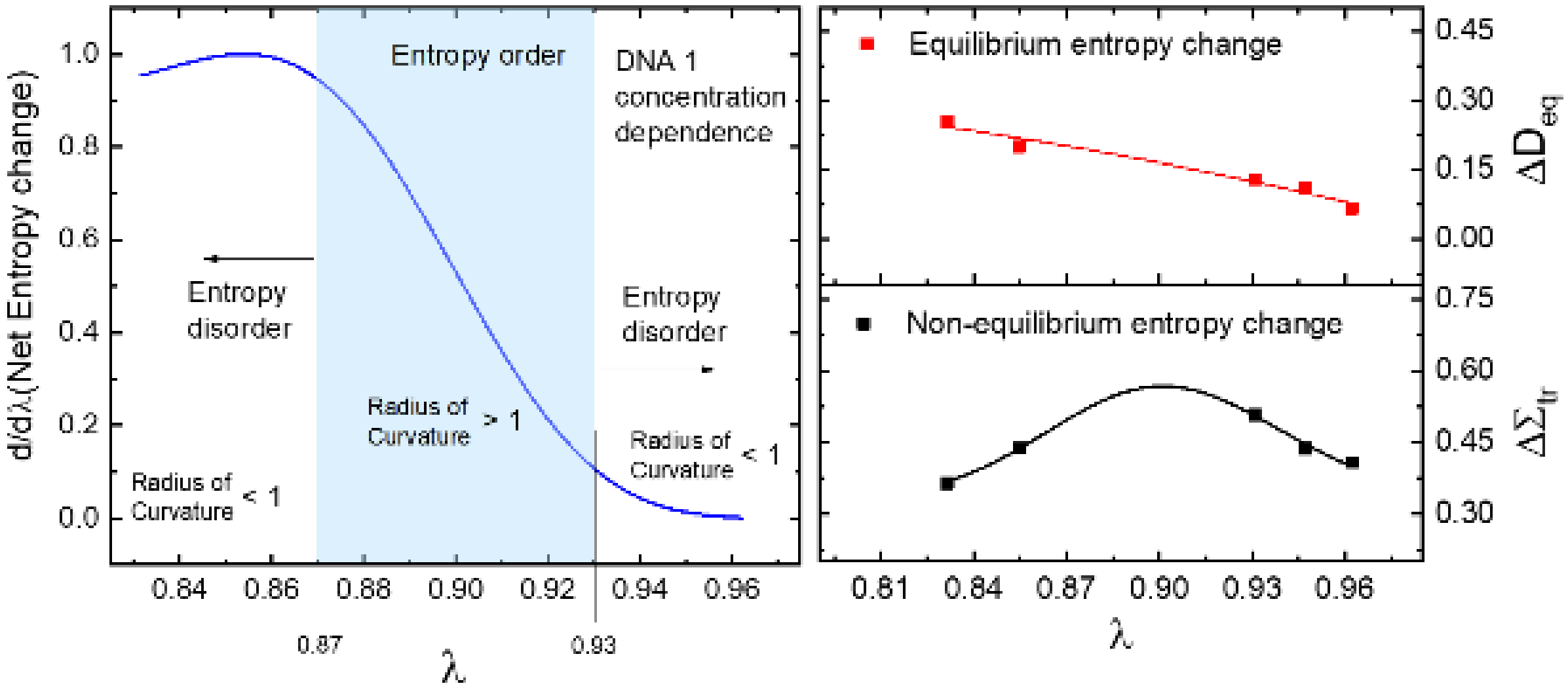

**Fig. S4.2|** Entropy rate change comparisons for equilibrium and non-equilibrium processes as a function of concentration of DNA 1 in solution. The equilibrium change can be approximated by a Boltzmann distribution. An apparent entropy maximization is exhibited as a function of the transitional ordering parameter $\lambda$ for the non-equilibrium change. The regime $\lambda > 0.87$ is entropy fluctuation dominated. Entropy ordering is favored where the radius of curvature of rate of entropy change is greater than 1. For curvatures less than 1, the transition is favored more towards to entropy disordering(*59*, *60*). For very low concentrations the molecules of the bath or the DNA solution start playing a greater role in the transition revealing a trend towards higher melting transition in comparison to nucleation transition.

**S5. Discussion on Entropy ordering vs disordering:**

The relation between order and entropy has perplexed scientists for a long time and has been discussed at length(*61–64*). Under equilibrium considerations one conjectures the satisfied condition of achieved largest disorder in somewhat of an ill-defined sense. Thus, entropy in general is often regarded as equivalent to disorder. This can be erroneous. Prigogine(*62*) discussed at length the notion of entropy ordering relevant to biological stable structures at equilibrium. Lansdberg(*61*) asked the relevant question - *Can entropy and "order" increase together?* Most generally entropy $S(n)$ can defined as

$$S(n(t)) = -k_B \sum_{i=1}^{n(t)} p_i \ln p_i \ , \tag{1}$$

where $p_i$ are the probabilities of the $n(t)$ degrees of freedom contributing to energy exchange, the degrees of freedom $n$ evolving in time $t$. Here $k_B$ is the usual notation of Boltzmann's constant. The notion of the time evolution of entropy was introduced by Kolmogorov and Sinai(*65*, *66*) and has also been considered at length for physical systems(*38*, *44*). The key to understanding entropy order is - it can be shown in actual situations concerning small systems that $S(n)$ can increase less rapidly with time in comparison to $k_B \ln n(t)$. Hence order can increase where the time rate of change of order $O(n)$ is given by

$$\frac{\frac{d}{dt}[O(n)]}{O(n)} = \frac{\frac{d}{dt}[n]}{n \ln n} - \frac{\frac{d}{dt}[S(n)]}{S(n)} . \tag{2}$$

Now $\frac{d}{dt}[n] \Big/ n \ln n \sim \frac{\tau}{\ln n}$, where $\tau$ is the measured timescale at which temperature inflections or irreversible energy exchange with the bath (DNA solution) is observed. Now typical $\tau \sim \exp(2.42) \approx 10$ as measured from experiments (Fig 2). The number of degrees of freedom that are involved in the fluctuation scale contributing to irreversible energy exchange with bath is $n \sim 8$, giving the first term in equation 2 above as $\tau / \ln n = (\exp(2.42)/\ln 8) \sim 5$. Essentially the degrees of freedom contribute to phase oscillations in the dynamic dissipation response as observed. The degrees of freedom are equivalent to Lyapunov exponents that are usually theoretically determined(*38*, *45*). Most generally speaking, here we are able to observe the same crudely from the phase oscillations in responses as shown below. On the other hand, the equilibrium entropy rate $\frac{d}{dt}\left[S(n)\right]/S(n)$ follows an intrinsic timescale $\tau_{res} = 2Q/f_{res}$ of the resonating microcantilver, giving an order $\frac{d}{dt}\left[S(n)\right]/S(n) \sim 0.8$ for the second term in equation 2. This is much smaller than the non-equilibrium rate order $\sim 5$ as derived above giving $O(n) > 1$. The equilibrium and non-equilibrium entropy rates as obtained from the measurements are shown below with same order of magnitude differences. Thus indeed, the non-equilibrium rate > the equilibrium rate leading to entropy order in the transition pathway.

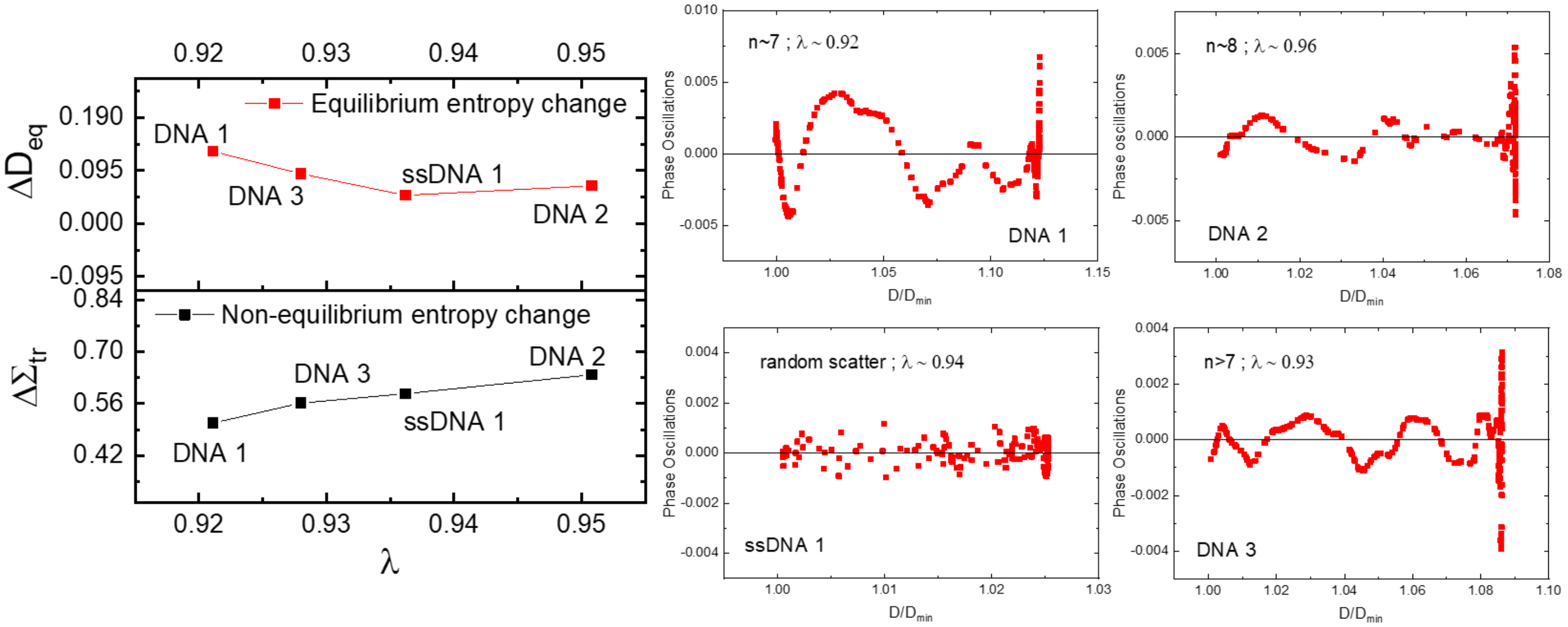

**Fig. S5|** Entropy rate change comparisons for equilibrium and non-equilibrium processes for different DNA oligomers.

The ordering parameter in ssDNA 1 can be explained from random generation of degrees of freedom $\sim n$ on $\pi - \pi$ dimer stacking or self hybridization forming beaded plasmid DNA conformations on substrates as shown before(*56*, *57*). Further work is being done to understand the nature of the phase oscillations and their relation to Lyapunov coefficients and specially the random phase excursions about 0 phase in ssDNA 1 that may have relevance to an apparent phase transition behaviour. These would be communicated in a later article not to digress from the main theme of the present article.

## S6. Experimental Methods:

### S6.1 dsDNA samples preparation

All the DNA samples were purchased from Integrated DNA Technologies (IDT, Coralville, IA, USA). Freeze-dried ssDNA samples were first dissolved in calculated amounts of phosphate buffer saline (PBS, Sigma-Aldrich, Oakville, ON, Canada) as $100\mu M$ stock solutions before further dilution. dsDNA samples were prepared by incubating the ssDNA's and their complimentary strands at room temperature with respective calculated concentrations.

### S6.2 Fabrication and Packaging

The microfluidic cantilever (MC516, Fourien Inc. AB, Canada) was fabricated using silicon nitride (SiN) as a structural material. To get consistent properties of deposition of the thin film of SiN, the deposition was performed using low pressure chemical vapor deposition (LPCVD)(*67*). The $500\mu m$ long (overhang length) and $57\mu m$ wide cantilever incorporates a microfluidic channel, which is $16\mu m$ in width and $980\mu m$ in overall length. Fig S6.1 left pane provides a top view of the microfluidic cantilever MC516. The U-shaped microfluidic channel is visible on the right inset. The left inset shows the topography of the microfluidic channel, showing the surface roughness of around 0.5 µm. A similar characteristic of the surface can also be observed in the right pane of Fig S6.1. The roughness comes from the crystallization of poly-silicon. This does not interfere in our dynamic dissipation measurements.

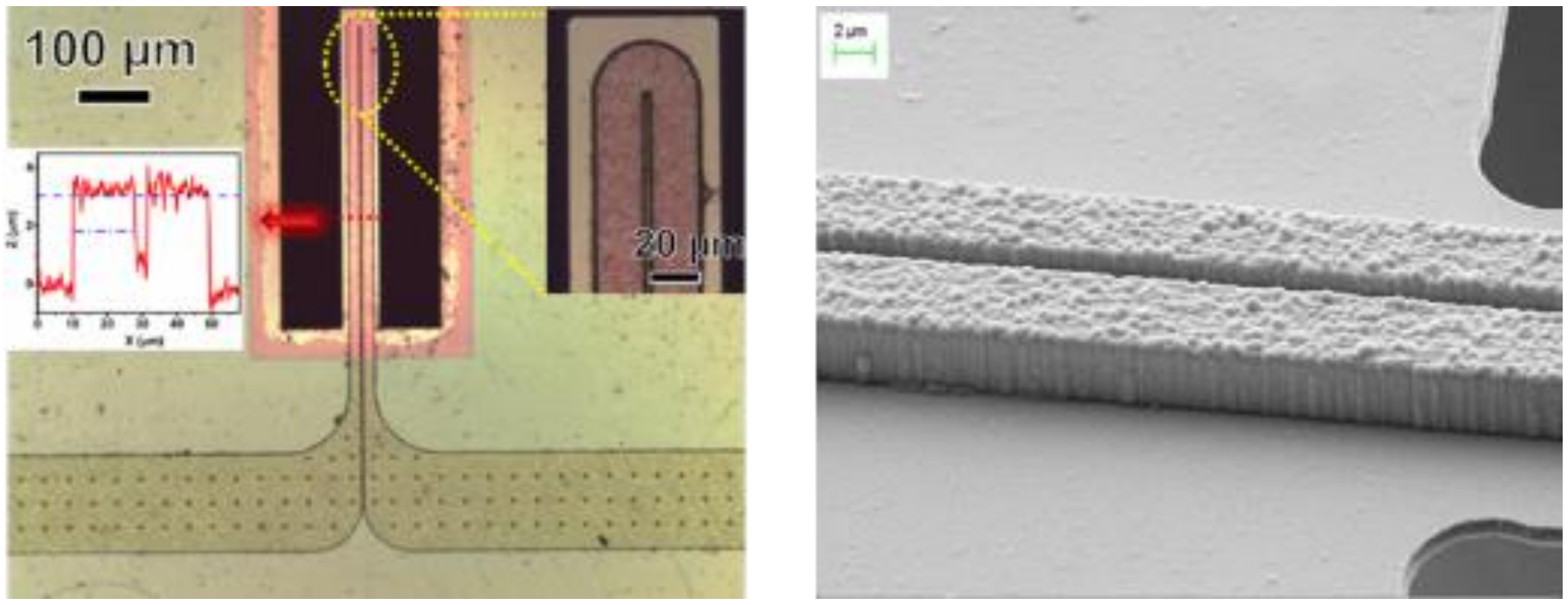

**Fig. S6.1|** SEM top view image (left pane) and white light confocal image (right pane) of a microfluidic cantilever, the channel size is 500×20×3 µm. The insertion on the SEM image is zoomed-in depiction of the tip of the cantilever. The confocal image highlights the microfluidic channel.

The microfluidic cantilevers are fabricated on $5mm \times 5mm$ silicon chips. These dimensions of the chips are optimum for ease of handling. The chips are equipped with two fluid inlet ports (width × length : $500\,\mu m \times 500\,\mu m$) on the bottom side. Using Buna-Nitrile O-rings, the microchips are placed in a chip holder made of Poly Acrylic acid (PAA). The smooth surface of acrylic provided a better adhesion of PDMS seal preventing evaporation of the fluids from the fluid inlet ports. In order to load a sample in the channel, a $5\mu L$ droplet is delivered at one of the inlet ports of the chip holder. Negative pressure is then applied from the other port of the chip holder. This creates vacuum inside the chip which helps in the streamlined loading of the liquid samples inside the channel. Once the presence of the liquid is confirmed by a change in the resonance frequency of the cantilever, the negative pressure is removed and a Polydimethylsiloxane (PDMS) membrane is used to seal off the injections ports, isolating the channels from the atmosphere.

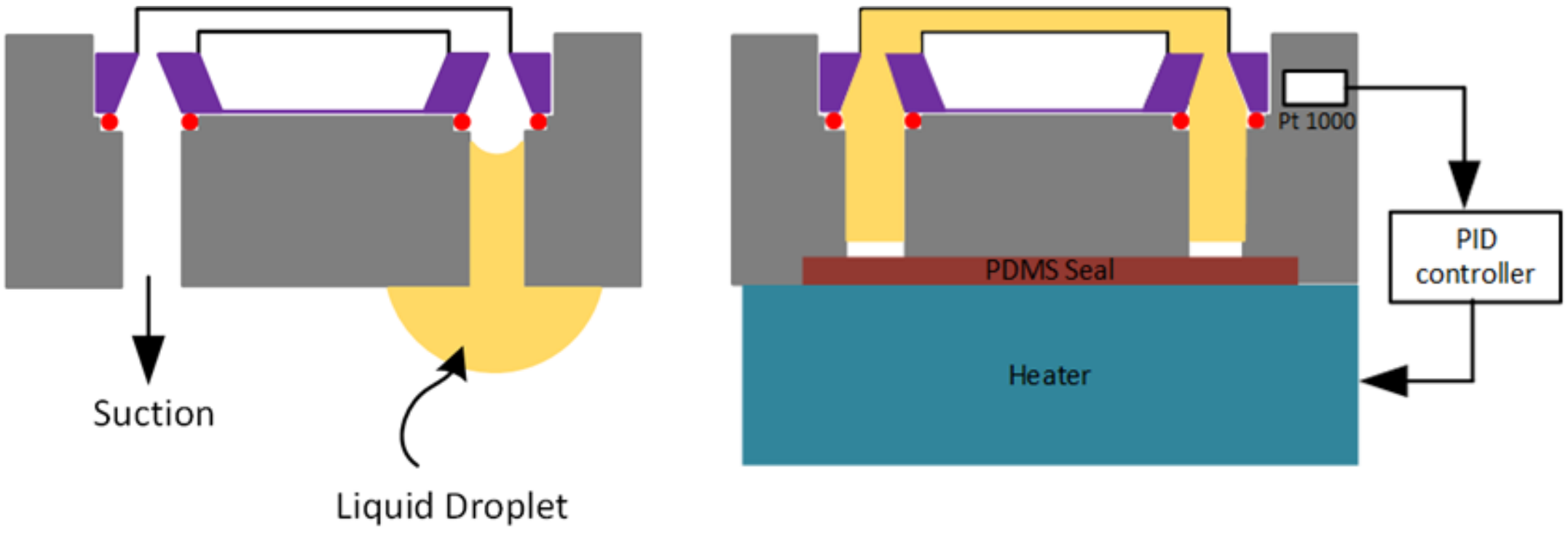

**Fig. S6.2|** Microfluidic packaging and sample loading schematic.

The DNA samples (confined in the cantilever) were heated using a $25W$, $5\Omega$ resistive heater (KAL25F, Stackpole Electronics Inc., Raleigh, NC, USA). The choice of the resistor provided the necessary flat and optimum area contact with the chip holder. The heating was controlled by a temperature controller (CN78020-C4, Omega Engineering Inc., Stamford, CT, USA). In a typical heating process, the temperature was ramped at a rate of $1^0C$ per minute by an optimized set of proportional-integral-differential (PID) parameters. The temperature profile was monitored and recorded by a Keithley 197 multimeter (Keithley Instruments, Cleveland, OH, USA) interfaced to PC with a LabView program. Once the heater was mounted on the chip holder, the PDMS seal was sandwiched between the heater and the holder as shown. This further provided a seal to prevent evaporation of the sample.

### S6.3 Q measurements

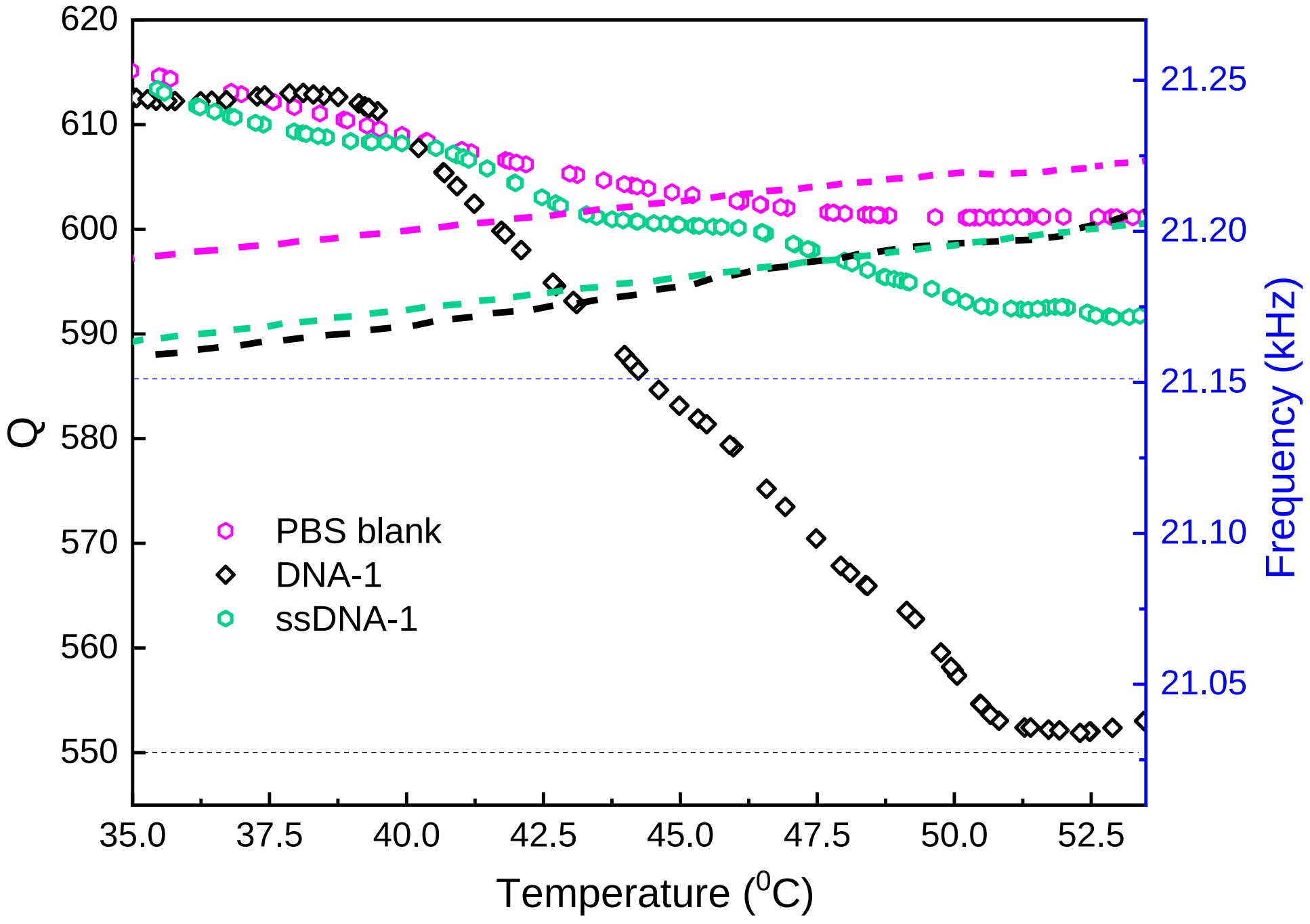

**Fig. S6.3|** Q factor decrease as a function of temperature, dsDNA shows a significant change while ssDNA shows limited change, PBS buffer is tested as blank reference: Q decrease with temperature ramping and frequency increase in the same range.

Miniature size of a microfluidic cantilever makes it an ideal platform for measuring changes in specific gravities of picolitres (pL) of confined liquids by continuous monitoring of its resonance frequency. In addition to quantifying the mass change (and the density) of the confined liquid sample, resonance response can also be used for monitoring the quality factor (Q) of the cantilever, a unit less parameter which describes the rate at which energy gets dissipated per oscillation. By monitoring the changes in the quality factor, it is possible to estimate the effect of changes in the viscosity of a sample present inside the cantilever. Most generally, Q can be defined as

$$Q = \frac{f_{res}}{\Delta f}, \quad (3)$$

where $f_{res}$ stands for the resonance frequency, $\Delta f$ is the full width at half maximum (FWHM). An increased viscosity of the liquid in the channel leads to higher energy dissipation with every resonant cycle, which causes a decrease in the Q-factor. In the DNA melting process, the dissociation of the dsDNA results in the dynamic viscosity change of the solution inside the channel.

The microfluidic cantilever (MC-516, Fourien Inc., Edmonton, AB, Canada) used for this work is $500\mu m$ in length, $20\mu m$ in channel width, $3\mu m$ in channel height. The structural material of the cantilever is silicon nitride which is quite compatible to bio-samples. The sensor chip was placed in a vacuum holder and operated at a $10^{-5} mTorr$ pressure in order to reduce air damping thus increasing the Q to desired levels for higher sensitivity. The resonance of the DNA filled cantilever was measured by a MSA-500 laser Doppler vibrometer (LDV Polytec, Irvine, CA, USA). The resonance frequency and Q were simultaneously measured. dsDNA sequences with variations in base pair numbers and/or composition (G-C ratio) were analyzed.

A typical dsDNA (DNA-1) melting in a microfluidic cantilever result shows that the Q decreases with increasing temperature (Fig S6.3). From 35 to $40^0 C$, the reduction in the Q is insignificant. However, the decrease in the Q is large from over 615 to about 550, in the temperature range of 40 to $50^0 C$, reflecting a transition behavior as described in detail in the article. After $50^0 C$, the decrease in the Q turns negligible again, indicating the end of the melting process. In comparison, from $35-50^0 C$, PBS blank sample shows a steady Q decrease from 615 to 602. Sample of ssDNA also shows a limited decrease in the Q from 614 to 590 over the temperature range mentioned. As illustrated in Fig S6.3, the frequency of the dsDNA, ssDNA and PBS all showed an increase with increasing temperature. These originates from the decrease in the density of liquid with increasing temperature. However, the observed resonance frequency deviations for all the samples were less than 2% at room temperature. These suggest that temperature dependent change of effective mass is small in our measurements.

The $T_m$ s were determined to be in the order of $47-51^0 C$ as shown in Fig 1. These values show very good agreement with the $T_m$ s determined from UV-vis $OD_{260}$ ($49^0 C$, see Fig S1.1). Nevertheless, compared to the UV-vis spectroscopy method, the sample volume used in the microfluidic cantilever sensor is only $100 pL$, which is much less than the $1 mL$ needed for the UV-vis measurement, making this a much sensitive technique.